\documentclass[lettersize,journal]{IEEEtran}
\usepackage{amsmath,amsfonts}
\usepackage{algorithmic}
\usepackage{algorithm}
\usepackage{array}
\usepackage[caption=false,font=footnotesize,labelfont=rm,textfont=rm]{subfig}
\usepackage{textcomp}
\usepackage{stfloats}
\usepackage{url}
\usepackage{verbatim}
\usepackage{graphicx}
\usepackage{cite}
\usepackage{bm}
\usepackage{multirow}
\usepackage[table]{xcolor}
\usepackage{color}
\usepackage{makecell}
\DeclareSubrefFormat{parens}{#1(#2)}

\usepackage{color}
\ifodd 1
  \newcommand{\com}[1]{\textbf{\color{red}(COMMENT: #1)}} %comment of the text
\else

  \newcommand{\com}[1]{}
\fi

\begin{document}

\title{Semantic Communication for Internet of Vehicles: A Multi-User Cooperative Approach}
% Multi-user semantic-aware architecture
% A Novel Paradigm for Multi-user Semantic-aware Networks
\author{Wenjun~Xu,~\IEEEmembership{Senior Member,~IEEE,}
Yimeng~Zhang,~\IEEEmembership{Graduate Student Member,~IEEE,}\\
Fengyu Wang,~\IEEEmembership{Member,~IEEE,}
Zhijin Qin,~\IEEEmembership{Senior Member,~IEEE,}
Chenyao Liu,
Ping Zhang,~\IEEEmembership{Fellow,~IEEE}

% \thanks{Wenjun Xu is with Key Lab of Universal Wireless Communications, Ministry of Education, Beijing University of Posts and Telecommunications, Beijing 100876, China, and also with Peng Cheng Laboratory,
% Shenzhen 518066, China (email: wjxu@bupt.edu.cn).

% Yimeng Zhang is with Key Lab of Universal Wireless Communications, Ministry of Education, Beijing University of Posts and Telecommunications, Beijing 100876, China (email: yimengzhang@bupt.edu.cn).

% Fengyu Wang is with School of Artificial Intelligence, Beijing University of Posts and Telecommunications, Beijing, China, 100876 (email:fengyu.wang@bupt.edu.cn)

% Zhijin Qin is with the School of Electronic Engineering and Computer Science, Queen Mary University of London, London E1 4NS, UK (email: z.qin@qmul.ac.uk)

% Chenyao Liu is with Key Lab of Universal Wireless Communications, Ministry of Education, Beijing University of Posts and Telecommunications, Beijing, China, 100876 (email: liuchenyao@bupt.edu.cn)

% Ping Zhang is with the State Key Laboratory of Networking and Switching Technology, Beijing University of Posts and Telecommunications, Beijing 100876, China, and also with Peng Cheng Laboratory,
% Shenzhen 518066, China (email: pzhang@bupt.edu.cn).

% (Corresponding author: Fengyu Wang).}% <-this % stops a space
}

% % The paper headers
% \markboth{Journal of \LaTeX\ Class Files,~Vol.~14, No.~8, August~2021}%
% {Shell \MakeLowercase{\textit{et al.}}: A Sample Article Using IEEEtran.cls for IEEE Journals}

% \IEEEpubid{0000--0000/00\$00.00~\copyright~2021 IEEE}
% % Remember, if you use this you must call \IEEEpubidadjcol in the second
% % column for its text to clear the IEEEpubid mark.

\maketitle

\begin{abstract}
Internet of Vehicles (IoV) is expected to become the central infrastructure to provide advanced services to connected vehicles and users for higher transportation efficiency and security.
A variety of emerging applications/services bring explosively growing demands for mobile data traffic between connected vehicles and roadside units (RSU), imposing the significant challenge of spectrum scarcity to IoV.
% In this context, an emerging problem for IoV is to deal with the spectrum scarcity due to the explosively growing demands for mobile data traffic required by new applications.
In this paper, we propose a cooperative semantic-aware architecture to convey essential semantics from collaborated users to servers for lowering the data traffic. 
{In contrast to current solutions that are mainly based on piling up highly complex signal processing techniques and multiple access capabilities in terms of syntactic communications, 
this paper puts forth the idea of semantic-aware content delivery in IoV.
Specifically, the successful transmission of essential semantics of the source data is pursued, rather than the accurate reception of symbols regardless of its meaning as in conventional syntactic communications.}
To assess the benefits of the proposed architecture, we provide a case study of the {image retrieval task for vehicles} in intelligent transportation systems. Simulation results demonstrate that the proposed architecture outperforms the existing solutions with fewer radio resources, especially in a low signal-to-noise-ratio (SNR) regime, which can shed light on the potential of the proposed architecture in extending the applications in extreme environments.
\end{abstract}

% \begin{IEEEkeywords}

% \end{IEEEkeywords}

\section{Introduction}
% 1. The challenges of IoV.
% \begin{itemize}
%     \item Data volume huge, limited communication resource
%     \item The importance of different semantics is different, need to allocate less resource to less important semantics; bit stream, treat each semantic instance equally.
%     \item Data of nearby RSUs are highly correlated, non-cooperative transmission leads to high interference and low efficiency
%     % \item IoV focuses on the integration of humans and vehicles, and human are an extension of a vechile's intelligence; however, the sensitivity to noise of M2M and M2H is distinct; need dedicated design for M2M and M2H separately; structure complex;
% \end{itemize}

% Figure 1: IoV overview: massive nodes
% \begin{figure*}
% \centering
% \includegraphics[scale=0.8]{IoV3.pdf}\\
% \caption{IoV overview: massive nodes.}\label{Fig_IoV}
% \end{figure*}

% Figure 2: Importance difference in semantic (e.g., image); 

% 2. Semantic communication is suitable for IoV
% \begin{itemize}
%     \item Extract semantics for tasks; save comm. resource; efficiency
%     \item Cooperatively learn the important semantics, allocate resource adaptively
%     \item reliability?
%     % \item Easy to extend to different tasks, structure concise;
% \end{itemize}

% % 1. the number of car dramatically increase (  ). so as the car accident. WHO, and financial; However, the deaths caused by car crash are in principle avoidable

Automobiles have become a daily necessity in modern society to provide a fast and convenient way to deliver goods and passengers. 
The rapid growth in the number of vehicles results in a dramatic increase in the time of traffic congestion, causing a waste of more than 56 hours and around 18 gallons of additional fuel for each commuter per year.
Moreover, according to the World Health Organization (WHO), approximately 1.35 million deaths and more than 20 million injuries are caused by road traffic crashes every year around the world \cite{driver}.
% Moreover, 

Aiming at improving the safety and efficiency of transportation system, the Internet of Vehicles (IoV) \cite{iov4_big_data} has been proposed, which enables information exchange among vehicles, users, and external infrastructures.
% The Internet of Vehicles (IoV) \cite{iov8_overview} has been proposed to improve the safety and efficiency of the transportation system
% With abundant information and integrating technologies of the Internet of Things (IoT) and Intelligent Transportation Systems (ITS), 
By integrating Internet of Things (IoT) and Intelligent Transportation Systems (ITS), 
IoV provides diversified services, such as traffic management, car navigation, and intelligent vehicle control, to avoid traffic accidents and ease traffic congestion.
% To support real-time applications\

To provide various real-time services to vehicular users, massive data should be transmitted to servers within corresponding delay restrictions while keeping the data integrity, leading to a huge demand for spectrum resources~\cite{iov4_big_data}.
% at the same time, requiring a large amount of bandwidth \cite{iov4_big_data}.
% Thus, IoV requires a large bandwidth to support its services.
% To meet the requirement, 
As a result, IoV enables multiple access capabilities, including cellular, WiFi, satellite, etc., for larger bandwidth.
Moreover, various advanced technologies, such as non-orthogonal multiple access (NOMA) and multiple-input multiple-output (MIMO), have been used to improve the spectrum efficiency.
However, the conventional communication systems have nearly approached the Shannon capacity limit. With the advent of more intelligent applications (e.g., autonomous driving), as well as the increasing number of vehicular users in IoV, the spectrum allocated for IoV hardly supports the big data transmission~\cite{iov4_big_data}.
% the spectrum sparsity is one of the biggest challenges for IoV.  a flood of infomation

\begin{figure*}
\centering
\includegraphics[scale=0.8]{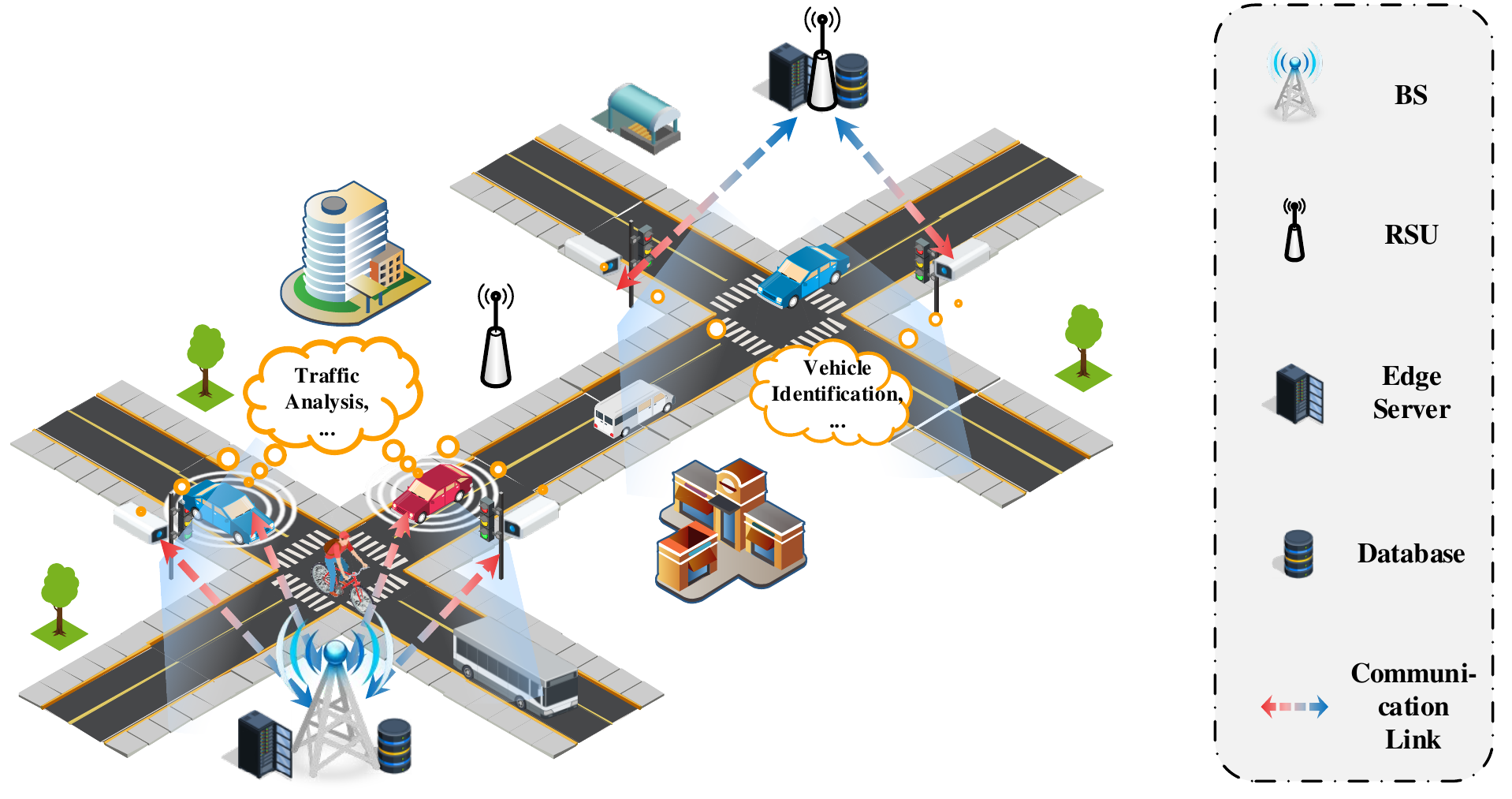}\\
\caption{The typical scenario of IoV: Correlated data of nearby users is transmitted to servers.}\label{Fig_iov}
\end{figure*}

On the other hand, many ITS applications, such as traffic congestion detection, autonomous driving, etc., require massive data from nearby vehicles and roadside units (RSU) \cite{arooj2021big}, as shown in Fig.~\ref{Fig_iov}. 
Although the collaboration of different data sources makes the performance better than stand-alone systems, the redundancy among the transmitted data \cite{iov2_redudancy} dramatically deteriorates the spectrum efficiency.
% Through collaboration of different data sources, the system outperforms the stand-alone system.
% However, current system 
Moreover, the raw data transmitted to servers may contain irrelevant information for specific tasks, leading to severe network congestion in IoV.

  {This paper aims to break the above limits by proposing a novel Cooperative Semantic Communication (Co-SC) architecture for IoV. The correlated semantic information from multiple users is extracted and transmitted via a shared channel, which is jointly recovered and exploited by the cooperative modules at the receiver for further processing.}
In particular, Co-SC extracts the intended ``meanings" or ``features" of the source data that are relevant to the transmission intention, and filters out the irrelevant and unessential information to lower the data traffic.
As a result, semantic communications\cite{zhang2021toward, qin2021semantic} can transmit less amount of data while preserving the effectiveness of communication, alleviating the transmission load significantly.
As one of the potential technologies for the sixth generation (6G) and beyond, semantic communications have drawn extensive attention in both academia and industry.
Preliminary works have shown the potential of semantic communication in improving both the transmission efficiency and reliability for supporting the end-to-end (E2E) transmission~\cite{xie2021deep,weng2021semantic, bourtsoulatze2019deep}. 
By jointly optimizing the semantic and channel coding, the point-to-point semantic transmission for text, image, and speech is achieved, outperforming the conventional syntactic-based system, especially in the low signal-to-noise (SNR) regime.
However, these works cannot be directly applied in multi-user scenarios in IoV. 

To deal with the multi-user scenario, our initial work~\cite{xie2021task} designs a multi-user semantic communication system for visual question answering (VQA) task, named MU-DeepSC. The correlated semantic information of users are incorporated at the receiver to get more accurate answer. 
However, the correlation between different users is only explored in the specific VQA task, and is not fully explored during the transmission.
% \textcolor{blue}{
To collaboratively utilize the correlated information of different users for more efficient transmission and intelligent tasks in IoV, in this work, we propose a general intelligent architecture, Co-SC, for multi-user applications in IoV.
% where the correlation among users is pre-learned and embedded in the structure of encoders and cooperative modules.
% }
The proposed Co-SC jointly designs the semantic encoder/decoder, where the redundant semantics of different users will be eliminated, meanwhile, the distinctive semantics of users that are relevant to the transmission goal will be extracted to improve the system performance.
Moreover, the correlation between the semantics of different users is further exploited in the cooperative joint source and channel (JSC) coding scheme. As a result, the decoder can better reconstruct semantic features of each user without extra transmission overhead, coping with wireless channel noise and impairment.

% The main contribution of this work is to propose a general intelligent framework for multi-user applications in IoV, where useful semantic features of different users will be jointly extracted and encoded to transmit to servers for various intelligent IoV applications. The proposed architecture offers many opportunities and open problem to the research community.

The remainder of this article is organized as follows.
We first present the framework and the functionality of each component of Co-SC.
Then, we implement a case study of an {image retrieval} application in IoV with the proposed Co-SC architecture, where extensive simulations have been conducted to investigate its effectiveness compared to the state-of-art baselines.
At the end of this article, future directions and concluding remarks are discussed. 

% , utilizing the proposed Co-SC architecture. Extensive simulation results have been conducted to investigate the performance of the proposed architecture.
% followed by a case study to show the effectiveness of the proposed framework. Then, future directions and open problems are discussed. Finally,  
% In the next sections, we first overview the proposed Co-SC framework, and 
% so that the noise and fading introduced by physical channel can be 
% Compared with traditional syntactic communication system, semantic communications\cite{sem4} aim to transmit the intended ``meanings" or ``features" of the source data that are relevant to the transmission intention, whilst those irrelevant and unessential information will be filtered out\cite{sem6}. Therefore, semantic communication systems can transmit smaller amounts of data while preserving the effective of communication. As a result, the data traffic would be reduced significantly.
% Moreover, since the semantic communication system can extract the content/intention of source data, the redundant information from other users can be effectively eliminated in CoSC, 

% where multiple agents/users with correlated source data transmit semantic features simultaneously to the server/receiver via a shared channel.

% \com{semantic information or semantic features; users or agents or terminals}

\section{Architecture of Cooperative Semantic Communications}
\begin{figure*}
\centering
\includegraphics[scale=0.42]{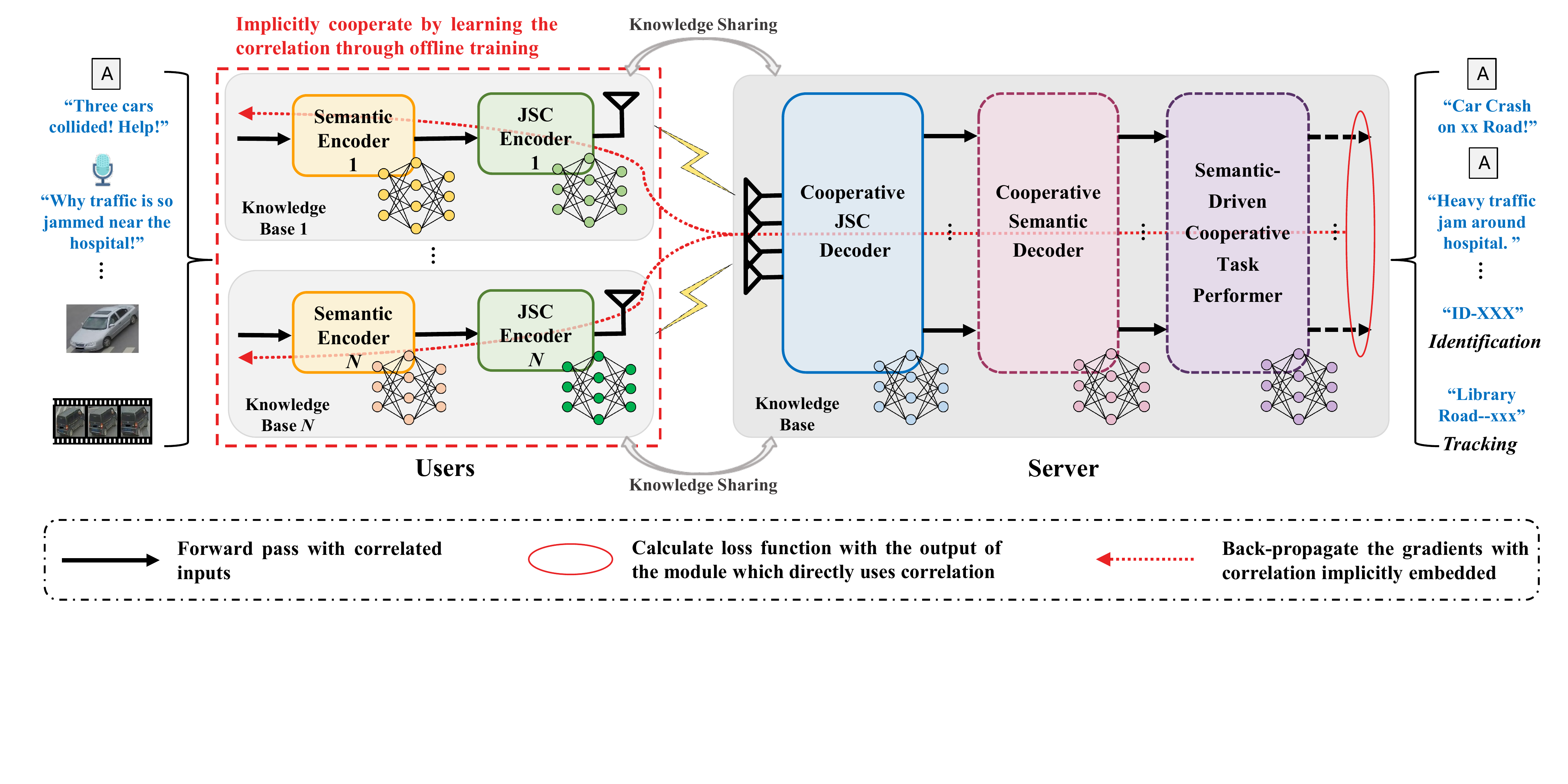}\\
\caption{The proposed architecture for general cooperative semantic communications.}\label{Fig_system}
\end{figure*}

In this section, we provide an overview of the proposed architecture Co-SC, which consists of semantic encoder/decoder (Sem-Codec), JSC encoder/decoder (JSC-Codec), and task-related modules, as shown in Fig.~\ref{Fig_system}. 
  {
Specifically, to achieve the required task, such as traffic analysis, pedestrian detection, vehicle tracking, etc., users/transmitters need to transmit correlated data to the server/receiver.
The correlation among users is pre-learned and embedded in the whole structure of Co-SC, including encoders at the transmitters and cooperative modules at the receiver.
At the transmitters, the essential semantic information is extracted by semantic encoders, and then, JSC encoders further encode the extracted semantic information to resist noise and interference during transmission.
}
At the server/receiver side, semantic features are recovered by the cooperative JSC decoder, and will be further processed by the cooperative semantic decoder and semantic-driven task performer on demand to fulfill the task at the receiver.
The detailed functionality of each component is listed in the following subsections. 
% The receiver will cooperatively decode the received signals and obtain the semantic features by the cooperative JSC decoder. The recovered semantic features will be further processed by the cooperative semantic decoder and semantic-based task performer on demands. 
% \textcolor{blue}{To provide data for the data-driven task at the server, agents encode the semantic information of source data with encoders. The server recovers the intended data with decoders and performs tasks with task-related modules on demands. The collected information helps the server to proceed other advanced tasks and make decisions about the subsequent data transmission.}  

\subsection{Semantic Encoder/Decoder}
% Question: Do we have semantic decoder? In the journal, do we need to add the semantic decoder in the architecture?
% \begin{itemize}
%     \item Semantic extraction is determined by different tasks
%     \item Decoder: recover the error in semantic level
% \end{itemize}
% Semantic encoder/decoder (SemCodec) is an essential part for SemCom.
  {
Generally, the semantic encoder is designed to extract the semantic information from the source data, which is a high-dimensional interpretation of the original data, emphasizing the meaning and goal-relevant part. Correspondingly, the semantic decoder recovers the source data or expresses it in other modalities from the high-dimensional semantic information according to the specific goals.}
% , which improves the compression efficiency compared with syntactic-based compression algorithms. With the semantic information, the semantic decoder recovers the source data or expresses it with other modal data according to the specific goals. 
For example, source images are {recovered} for the data-reconstruction-oriented system~\cite{bourtsoulatze2019deep}, and speech signals are reconstructed as the text transcriptions for speech recognition task~\cite{qin2021semantic}.
% while for the speech recognition task, semantic decoder generates the text transcriptions of speech signals~\cite{weng2021semantict}. 
% The Semantic encoder and decoder are optimized jointly to ensure that the extracted semantic information is efficient and robust for recovery. The optimization objective is determined based on the transmission goals.

Inspired by the semantic-level correlation among users, Co-SC incorporates a cooperative semantic decoder to leverage the semantic-level correlation among users, as shown in Fig.~\ref{Fig_system}. Compared to the E2E design for the single transceiver~\cite{xie2021deep,weng2021semantic, bourtsoulatze2019deep}, the advantages of cooperative design are twofold.
% \textcolor{blue}{
% First, by jointly optimizing the semantic encoders of multiple users and the cooperative semantic decoder, the correlation among users can be learned by the Sem-Codec.
% As a result, the semantic encoder is capable of
% % implicitly cooperating, 
% extracting informative semantic information of each user, while the redundancy can be compressed to improve the compressing efficiency.
% }
% Specifically, during training, the loss function is calculated with the outputs of the model, which are obtained by cooperatively processing the correlated semantic information. The gradients with correlation information implicitly embedded are back propagated through the whole model with back propagation (BP) algorithm, jointly optimizing the decoder and the encoders of multiple users.
% With back propagation (BP) algorithm, the gradients 
First, by jointly optimizing the semantic encoders and the cooperative semantic decoder, the correlation among users can be learned by the Sem-Codec.
As a result, distinctive semantic information of each user can be obtained by semantic encoders, while the redundancy can be compressed to improve the compressing efficiency.  
Second, the inherent correlation among users can be implicitly used for the error correction at the semantic level, further improving the accuracy of Sem-Codec.

Note that for some intelligent tasks, such as machine-to-machine (M2M) applications, semantic features can be directly used by the semantic-driven task performer for intelligent tasks. In such cases, the semantic decoder can be omitted, however, the correlation among users can still be learned by semantic encoders by jointly optimizing the cooperative task performer and semantic encoders with back propagation (BP) algorithm.
% Second, by jointly optimizing the semantic encoders and the cooperative semantic decoder, the correlation among users can be learned by the SemCodec. Therefore, more distinctive semantic information

% Inspired by the semantic-level correlation among users, we design a cooperative semantic decoder in Co-SC and jointly optimize it with the semantic encoders of multiple users. 
% The advantages are twofold. 
% First, the correlation among users can function like the error correction at the semantic level, further improving the accuracy of semantic decoding. 
% Meanwhile, compared with decoding separately, the cooperation at the semantic decoder provides the semantic encoders a more direct access to the correlation among users and facilitates a more efficient encoding scheme. 
% Specifically, by end-to-end training with a large set of data, semantic decoder can learn the correlation among users, which also makes difference to semantic encoders by back propagation (BP) algorithm, guiding them to extract more distinctive semantic information and compress the redundant information. 
% Note that the semantic decoder is one of the optional modules in Co-SC. When serving some kind of intelligent tasks, where the recovered semantic information can be used for tasks directly, the semantic decoder can be omitted.

\subsection{Joint Source and Channel (JSC) Encoder/Decoder}
% JSC-Codec is used for combating channel distortion.
The functionality of JSC-Codec is to resist channel distortions.
In specific, the JSC encoder is applied to encode the extracted semantic information as channel input symbols, while the JSC decoder recovers the semantic information with the received noisy symbols.
Unlike the JSC coding scheme in conventional communication systems, where channel symbols are obtained regardless of the transmission meaning, in Co-SC, JSC-Codec operates at the semantic level, where the channel symbols are obtained with the awareness of semantic information.
% As a result, the important semantics can be protected with more channel symbols to 
% Different from the existing source coding and channel coding algorithms with single function, the JSC encoder works more comprehensively, which compresses the semantic information to fit the the channel dimension and at the same time, enhances the robustness of the semantic information to channel distortion. 

Specifically, semantics with different importance levels are protected with a different number of symbols to enhance the robustness of the semantic information to channel distortions implicitly.
Moreover, to further leverage the semantic-level correlation among users, in Co-SC, a cooperative JSC decoder is designed to recover the transmitted semantics of multiple users jointly, as shown in Fig.~\ref{Fig_system}.
% Compared to the separated design in previous point-to-point (P2P) semantic communication system, 
Note that the semantic-level correlation narrows the scope of potential symbols, and the received symbols of users can provide a reference for each other, as a result, higher accuracy of JSC-Codec can be achieved with the cooperative JSC decoder.

% , so that the uncertainty of decoding is reduced.

% In Co-SC, we design a cooperative JSC decoder to recover the transmitted semantic information for multiple users. Semantic-level correlation among users is also helpful in combating channel distortion. 
% In specific, the uncertainty of decoding is reduced since correlation narrows the scope of potential symbols.
% Furthermore, the received symbols of users can provide reference for the recovery of each other.
% By jointly training the JSC encoders of multiple users and the cooperative JSC decoder, the optimal coding and decoding strategies are learned.
% There are many options for the optimization objective, including mean square error (MSE) loss function, mean absolute error (MAE) loss function, etc.. 

\subsection{Semantic-Driven Cooperative Task Performer}
The semantic-driven task performer is used to achieve specific tasks with the recovered semantic information from multiple users
% , such as traffic analysis, pedestrian detection, vehicle tracking, etc.}
The structure of the semantic-driven task performer adapts to specific tasks.
For example, convolution neural networks (CNN) are generally used for image-based tasks, and long-short-term memory (LSTM) is widely used for speech recognition, etc..
Note that for the tasks oriented toward information recovery, the task performer can be omitted, where 
the output of the semantic decoder may directly achieve the intelligent goals.

% In Co-SC, the semantic-level correlation among users is considered and we design the semantic-driven cooperative task performer, where the task is cooperatively performed by combining the information provided by users. 
In Co-SC, the semantic-level correlation and distinctions among users are leveraged by the semantic-driven cooperative task performer, and the task is cooperatively performed by combining the information provided by distinct users. 
The combination way can be adaptively designed according to the type of task and correlation.
For example, for the tasks with partially correlated semantic information, the semantic-driven cooperative task performer can be designed as two cascade networks, where
the correlation with the recovered semantic information of each user is captured in the first network, and then, the second network combines the correlated information into a global feature and concatenates it with the distinctive semantic information of different users as an enhance semantic feature for the task. 
The enhanced semantic information can facilitate better task performance in such a scenario.

% The task performer is the other optional module in Co-SC and is implemented for those intelligent tasks, which can be performed with the recovered semantic information directly, such as object classification, object identification, etc..
% We design a cooperative task performer to incorporate the semantic information of multiple users for better task performance.
% Benefiting from the semantic-level correlation among users, the individual semantic information can all provide effective information for the common goal, which makes the cooperation feasible and efficient.

\subsection{Knowledge Base}
The knowledge base is the basis of semantic communications and is one of the sources of the subjectivity of semantic information. Just like human beings, the way a person recognizing and depicting the world is determined by his knowledge learned and accumulated in his life, which differs from person to person. In semantic communications, given a goal, users first analyze and understand it based on their background knowledge and then perform coding or decoding. Semantic-level coding can be interpreted as the process of extracting and encoding the goal-related semantic information from source data, while semantic-level decoding can be considered as interpreting semantic information in the modal required by the transmission goal. Hence, the difference of background knowledge will deteriorate the performance of semantic communications severely. 
In general, before data transmission, transceivers will share their knowledge
% by exchanging knowledge directly or with a shared knowledge base at a central server through specific links. 
by acquiring or exchanging knowledge with a shared knowledge base at a central server through a specific link.
% In general, before data transmission, transceivers will share their knowledge by acquiring or exchanging knowledge with a shared knowledge base at a central server through a specific link. 

{In Co-SC, we assume that the background knowledge is already shared among users and the server. This can be achieved by jointly training the whole neural network offline with the common-accessed dataset, equipping the encoders and the decoder with the same cognition to the given transmission goal. 
Note that research about how to achieve efficient global knowledge sharing for multi-user semantic communications or semantic networks is a particular research topic, which is out of the scope of the manuscript.}
% Note that the JSC encoder and cooperative JSC decoder are also designed based on background knowledge since semantic-aware coding/decoding is performed by lay more resources on semantic information of high significance for protection.

% Or We have
% \section{Architecture of Cooperative Semantic Communications}
% Figure 3: frame work/ system architecture for general cooperative semantic communications; 
% \subsection{Components at RSU side}
% \subsubsection{semantic encoder}
% Related with tasks
% \subsubsection{channel encoder}
% Important semantics need more protection?
% \subsection{Components at Server side}
% \subsubsection{Joint channel decoder}
% Correlation of the semantics for different users is helpful in combating channel distortion 
% % denoise;

% \subsubsection{semantic decoder}
% optional? kind of recover the error in semantic level
% \subsubsection{Semantic fusion/correlation}
% optional
% \subsubsection{Identifier/Classifier?}
% semantic feature
% The relationship of the components at the receiver side is not that clear

% % Figure of Multi-task architecture as the proof of easy to extent

% % \section{Proposed Cooperative semantic communication system}

\section{Case study: Cooperative ID-retrieval in IoV}
\begin{figure}
\centering
\includegraphics[scale=0.7]{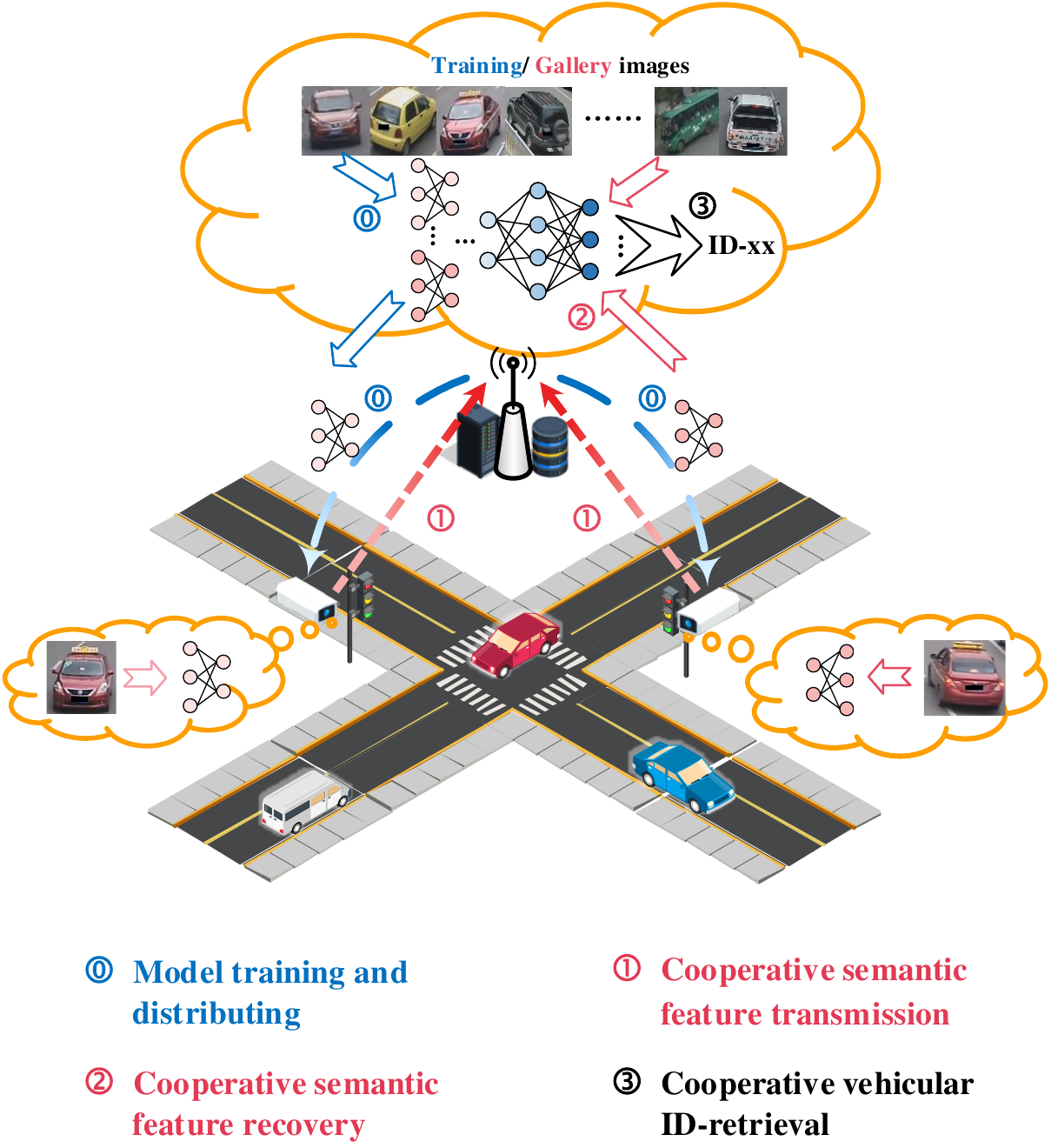}\\
\caption{The framework for the cooperative vehicular ID-retrieval task.}\label{Fig_reid_process}
\end{figure}

In this section, the proposed Co-SC is implemented to support IoV, in which the image retrieval task is essential~\cite{yimeng2022}. To provide data support for the intelligent tasks at the central server, such as suspicious vehicle positioning, cameras at the RSUs need to transmit the captured images to the server, as shown in Fig.~\ref{Fig_reid_process}.
The server retrieves the identifications of the received images (\textbf{\textsl{a.k.a.}} query images) by calculating the distance between their semantic features and that of the gallery images, which are only accessible at the server.
As shown in Fig~\ref{Fig_reid_process}, the nearby cameras tend to capture the images of the same vehicle from different angles, which results in the semantic-level correlation and thus, enables the cooperative semantic communication and identification.
Utilizing the Co-SC architecture for this task, only the semantic features of images are cooperatively transmitted to the server, instead of entire images, for the task without image reconstruction.
The entire Co-SC based multi-user system is achieved by the deep-leaning-based (DL) method, where the DL modules are trained offline at the server, and the trained models of transmitters will be implemented for cameras before transmission, as shown in Fig~\ref{Fig_reid_process}. Note that the model training and distributing only need to be performed once unless the distribution of the collected data changes dramatically.
% \textcolor{blue}{A deep-leaning-based (DL) task-oriented multi-user semantic communication system is designed, which is trained offline at the server and the trained models of transmitters will be implemented for cameras before transmission.}
% For this task, we design a task-oriented cooperative semantic communication system, where 

% Figure 4: System architecture
\begin{figure*}
\centering
\includegraphics[scale=0.52]{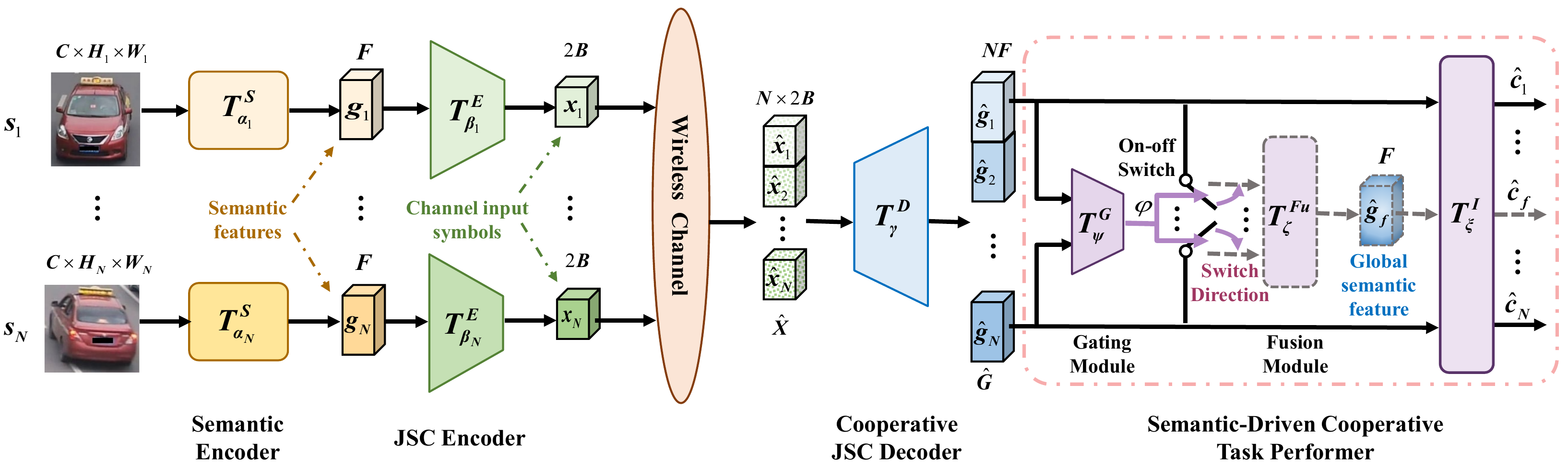}\\
\caption{The neural network structure of the proposed Co-SC for image retrieval task of vehicles. $\bm{T}$ is the transfer function of a module and the subscript denotes the trainable parameter set.}\label{Fig_proposed_architecture_DeepSC-COI}
\end{figure*}

\subsection{Methodology}

We consider the uplink scenario, where $N$ single-antenna cameras simultaneously transmit data via a shared channel to the central server equipped with $M$ antennas.
The detail implementation of the neural network structure of Co-SC for this ID-retrieval task is shown in Fig.~\ref{Fig_proposed_architecture_DeepSC-COI}. 

\textbf{Cooperative semantic feature transmission.} At the transmitter, given images $\bm{s}_i \in \mathbb{R}^{C \times H_i \times W_i}, i=1, ..., N$, cameras first extract the semantic features of images $\bm{g}_i \in \mathbb{R}^F, i=1, ..., N$ with the semantic encoder, where $W_i$, $H_i$, $C$ are the width, height, number of channels of images and $F$ is the dimension of semantic features.
Then the JSC encoder maps the semantic features to the channel input symbols $\bm{x}_i \in \mathbb{R}^ {2B}, i=1, ..., N$, where $B$ is the number of the transmitted complex symbols and $2B$ is the result of the transformation to real-valued symbols when processing.
  {Note that the complex channel input symbols can be considered as the counterpart of the symbols of conventional modulation.}
Average power constraint is used to normalize the channel input symbols before they are transmitted and all cameras are constrained with the same average power $P$.
The wireless channel can also be modeled as a layer with non-trainable parameters, as long as the channel transfer function is differentiable. For the non-differentiable cases, the channel can be approximated with a generative adversarial net (GAN)~\cite{ye2020deep}.

\textbf{Cooperative semantic feature recovery.} At the receiver, perfect channel state information (CSI) is assumed for signal detection. The detected symbols $\bm{\hat{X}} \in \mathbb{R}^ {N \times 2B}$ are fed into the cooperative JSC decoder, which outputs the concatenated semantic features of multiple cameras, represented as $\bm{\hat{G}} \in \mathbb{R}^ {NF}$. For the ID-retrieval task, the recovered semantic features can be directly used for identification and hence, the cooperative semantic decoder is not involved in this case. 

\textbf{Cooperative ID-retrieval.} The cooperative task performer is specifically designed for the vehicular ID-retrieval task. To incorporate the semantic features of multiple cameras, a fusion module is applied to fuse the recovered individual semantic features as a global semantic feature $\bm{\hat{g}}_f \in \mathbb{R}^F$.
% Dynamical-tailored weight allocation strategy is learned for fusion by training with a large set of images, which lays more weight on the semantic features of higher effectiveness level and can contribute more to the identification task.
Dynamical-tailored weight allocation strategy is learned for fusion by training with a large set of images, where {higher weights are} assigned to the semantic features of higher effectiveness to contribute more to the identification task.
% which lays more weight on the semantic features of higher effectiveness level and can contribute more to the identification task.
Finally, the identifier retrieves the identification with the global semantic feature. The identification corresponding to the source image is indicated by the maxima of probability vector $\bm{\hat{c}}_f \in \mathbb{R}^S$, where $S$ is the total number of identifications of training data. Note that the identifier is only used during training to learn efficient feature representations, while in the testing stage, the task is performed by calculating distances between semantic features.

% To adapt to the realistic application, a gating module is implemented to verify whether the images of cameras are correlated at the semantic level, based on which different identification strategy is used.
To make the system more robust in practical applications, a gating module is implemented to verify whether the images of cameras describe the same vehicle, based on which different identification strategies are used.
% The input of the gating module is the subtraction of every two cameras' recovered individual semantic features,
The input of the gating module is the subtraction of the recovered semantic features between two individual cameras,
while the output is a binary indicator {$\varphi$, where zero value} means that two semantic features are not of the same identification and vice versa.
% The individual semantic features of the same identification will be inputted to the fusion module and the identification of their images will be retrieved by calculating the distance between the global semantic features. For the remaining semantic features, the task will be performed separately without fusion.
The semantic features of the same identification will be fed into the fusion module to get the global semantic features for vehicular ID-retrieval.
For the remaining semantic features describing distinct vehicles, the task will be performed separately without fusion.
The detailed implementation of each module is listed in Table~\ref{parameter setting of proposed}.

% and the identification of their images will be retrieved by calculating the distance between the global semantic features. For the remaining semantic features, the task will be performed separately without fusion.

\subsection{Training Strategy}
The proposed system is trained with a four-stage strategy at the server. We first train the whole network without the gating module, referred as the backbone network, and then the gating module is trained with other modules frozen.
% \textcolor{blue}{All the stages are trained until the loss function converges.}
% A four-stage strategy is adopted.

First, the semantic encoder and the identifier are trained to learn the feature extraction strategy without being attached to other modules. 
% The loss function is the combination of cross-entropy (CE) loss function and triplet loss function with hard example mining (TriHard) loss function, which is commonly used in image retrieval task to obtain robust representation of semantic features~\cite{Luo_2019_CVPR_Workshops}.
% The loss function of this stage is represented as
% \begin{equation}
% \begin{aligned}
% \mathcal{L}_1 = \mathcal{L}^{\rm T}\left(\bm{g}\right) + \mathcal{L}^{\rm{CE}}\left(\bm{c}, \bm{\hat{c}}\right).
% \label{Eq_L1}
% \end{aligned}
% \end{equation}
This stage only needs to be performed once, and the trained semantic encoder will be loaded for individual cameras as the pre-trained model. The trained identifier is loaded at the receiver and is shared by all cooperative cameras.
In the second stage, the JSC encoders of multiple users and the cooperative JSC decoder are jointly trained to minimize the distance between the recovered semantic features and the transmitted ones, which are extracted by the trained semantic encoders.
  {
The distance is measured by the mean-squared error (MSE) in this case study. }
The parameters of other modules will not be updated in this stage. 
% The second stage focuses on the JSC encoders and the cooperative JSC decoder. 
% Aiming at the recovery accuracy of semantic features, the mean-squared-error (MSE) loss function is used to measure the distance between the recovered semantic features and the transmitted ones, which are extracted by the trained semantic encoders.
% To achieve the optimum performance while ensuring fairness,
% % To achieve the global and fair optimization, 
% the JSC encoders of multiple cameras are jointly trained, and a dynamically-weighted sum of MSE loss function is adopted.
% which is represented as
% \begin{equation}
% \begin{aligned}
% \mathcal{L}_2 & = \mathcal{L}^{\rm{MSE}}\left(\bm{g}_1, ...,\bm{g}_N, \bm{\hat{g}}_1, ..., \bm{\hat{g}}_N\right)\\
% & = \lambda^p_1\mathcal{L}^{\rm{MSE}}_1\left(\bm{g}_1, \bm{\hat{g}}_1\right) + \cdots + \lambda^p_N\mathcal{L}^{\rm{MSE}}_N\left(\bm{g}_N, \bm{\hat{g}}_N\right),
% \label{Eq_L2}
% \end{aligned}
% \end{equation}
% where $\lambda^p_i$ is the weight of the $i$-th camera $i$ at $p$-th mini-batch. 
% The weights of cameras are first initialized as $1/N$ and updated according to the proportion of their MSE loss function in the total MSE loss. The parameters of other modules will not be updated in this stage. 
In the third stage, the whole backbone network is jointly trained.
% with the combination of CE loss functions, which consists of the global CE loss function and the weighted sum of individual CE loss functions. The former is calculated with the identification result obtained according to the fused global semantic feature, while the latter is calculated with individual identification results. 
% The loss function of the third stage is represented as
% \begin{equation}
% \begin{aligned}
% \mathcal{L}_3 = & \mathcal{L}^{\rm{CE}}\left(\bm{c}_f, \bm{c}_1, ..., \bm{c}_N, \bm{\hat{c}}_f, \bm{\hat{c}}_1, ..., \bm{\hat{c}}_N\right) = \underbrace{\mathcal{L}^{\rm{CE}}\left(\bm{c}_f, \bm{\hat{c}}_f\right)}_{\rm{Global~CE}} +\\
% & \underbrace{{\mu}_1^p \mathcal{L}^{\rm{CE}}\left(\bm{c}_1, \bm{\hat{c}}_1\right) + \cdots + {\mu}_N^p \mathcal{L}^{\rm{CE}}\left(\bm{c}_N, \bm{\hat{c}}_N\right)}_{\rm{Individual~CEs}},
% \label{Eq_L3}
% \end{aligned}
% \end{equation}
% where $\mu^p_i$ is the weight of camera $i$. 
% The same update mechanism as the second stage is adopted for individual CE loss functions, which are involved to guarantee the effectiveness of the individual semantic features provided for the fusion module.
The gating module is trained in the final stage to evaluate the correlation based on the individual semantic features recovered by the trained backbone network obtained through the previous three stages.
% \textcolor{blue}{Except the first stage, the inputs of the other stages are all correlated, which means that the model is trained when multiple users transmit the images of the same vehicle. }

% The overall system is trained with CE loss function to minimize the difference between the distribution of predicted results and ground truths. 
% The objective is CE loss function, which is used to minimize the difference between the distribution of predicted results and ground truths. 

% After the backbone network is trained, the gating module is trained to verify the correlation based on the individual semantic features recovered by the trained backbone network. The objective is CE loss function, which is used to minimize the difference between the distribution of predicted results and ground truths. 

\section{Simulation Results}
This section presents the evaluation results of the proposed Co-SC on the VeRi-776 dataset~\cite{liu2016large}. Euclidean distance is applied to measure the distance between semantic features, one of the most widely used methods in image retrieval tasks. The calculated distances are ranked as a list in ascending order. Based on the list, two of the most popular performance metrics, rank-$n$ accuracy and mean average precision (mAP), are evaluated. 
% \textcolor{blue}{Specifically, rank-$1$ accuracy indicates the proportion of the query images, that are correctly retrieved by the first gallery image in the list.
% mAP reveals the degree that all the accurately retrieved gallery images rank high in the candidate list, providing a more comprehensive evaluation of the retrieval task.}
Specifically, rank-$n$ accuracy indicates the proportion of query images that are correctly retrieved by the first $n$ results in the list.
In the following, rank-1 accuracy is used to assess the performance.
% Here we use rank-1 accuracy. 
mAP measures the mean of average precision (AP), indicating the proportion of the correctly retrieved gallery images in the gallery image set.

% During training, the input images are first resized to $256 \times 256$ and augmented. 
% Stochastic gradient descent (SGD) optimizer with a momentum of 0.9 is used, and cosine learning rate decay is applied to update the learning rate.

The proposed Co-SC scheme is compared with traditional transmission methods and DL-based {semantic transmission} methods with different levels of cooperation as baselines.
  {The evaluation is performed with both the correlated and uncorrelated cases (i.e., cameras capture different vehicles) in the test data.}
% the input images of multiple users are not always correlated. We randomly set a part of the input images to be of different vehicles.
The implementation details of the baselines are listed in the following.
% We compare the proposed Co-SC-based method with two baselines, i.e., traditional transmission methods and deep-learning-based (DL) methods.
\begin{itemize}
\item   {\textbf{Traditional transmission methods}: The images are encoded and reconstructed with traditional algorithms. 
% ResNet-50 pretrained on the VeRi-776 dataset is used to extract features from reconstructed images. The following (non-cooperative) baselines are implemented to evaluate the effectiveness of semantic communications.
}
\begin{itemize}
\item[-]   {\textbf{Digital transmission method}: 
% from which the semantic features are extracted by ResNet-50 pretrained on VeRi-776 dataset. 
JPEG, LDPC with $3/4$ rate and BPSK are used for source coding, channel coding, and modulation, respectively. This method is referred to as \textit{JPEG+LDPC+BPSK} in the simulation results.}

\item[-]   {\textbf{Analog transmission method}: SoftCast~\cite{softcast1} performs analog image transmission, where the DCT coefficients of images are directly transmitted after power scaling and modulation. }
\end{itemize}

\item \textbf{DL-based baselines}: The architecture of the transmitter is the same as that of the proposed Co-SC scheme, while the receiver differs. Two architectures are investigated to demonstrate the effect of cooperation with different levels. 
\begin{itemize}
\item[-]   {\textbf{\textit{Co-SC w/o fusion}}: The individual semantic features are recovered with the cooperative JSC decoder, while the fusion module is not adopted at the task performer.
This method neglects the cooperation at the semantic-driven task performer (the fusion module in this case study).
% This method is with partial cooperation and is to evaluate the functionality of the semantic fusion module.
}

\item[-]   {\textbf{\textit{DL-S}}: The individual semantic features are recovered with separate JSC decoders, which is adopted in~\cite{xie2021task}, instead of the cooperative JSC decoder. DL-S neglects the cooperation at both the JSC decoder and the semantic-driven task performer.
% This method is non-cooperative and is to evaluate the functionality of the cooperative JSC coding scheme.
}
\end{itemize}
Note that for both DL-based baselines, the task is performed individually with the recovered semantic features $\bm{\hat{g}}_i$ of each camera. All the cameras share the parameters of the identifier during training. The same training strategy as Co-SC is adopted. 
\end{itemize}

\begin{figure*}[tbp]
	\centering
	\subfloat[Retrieved results.]{\label{Fig_retrieval}\includegraphics[width=9.35cm]{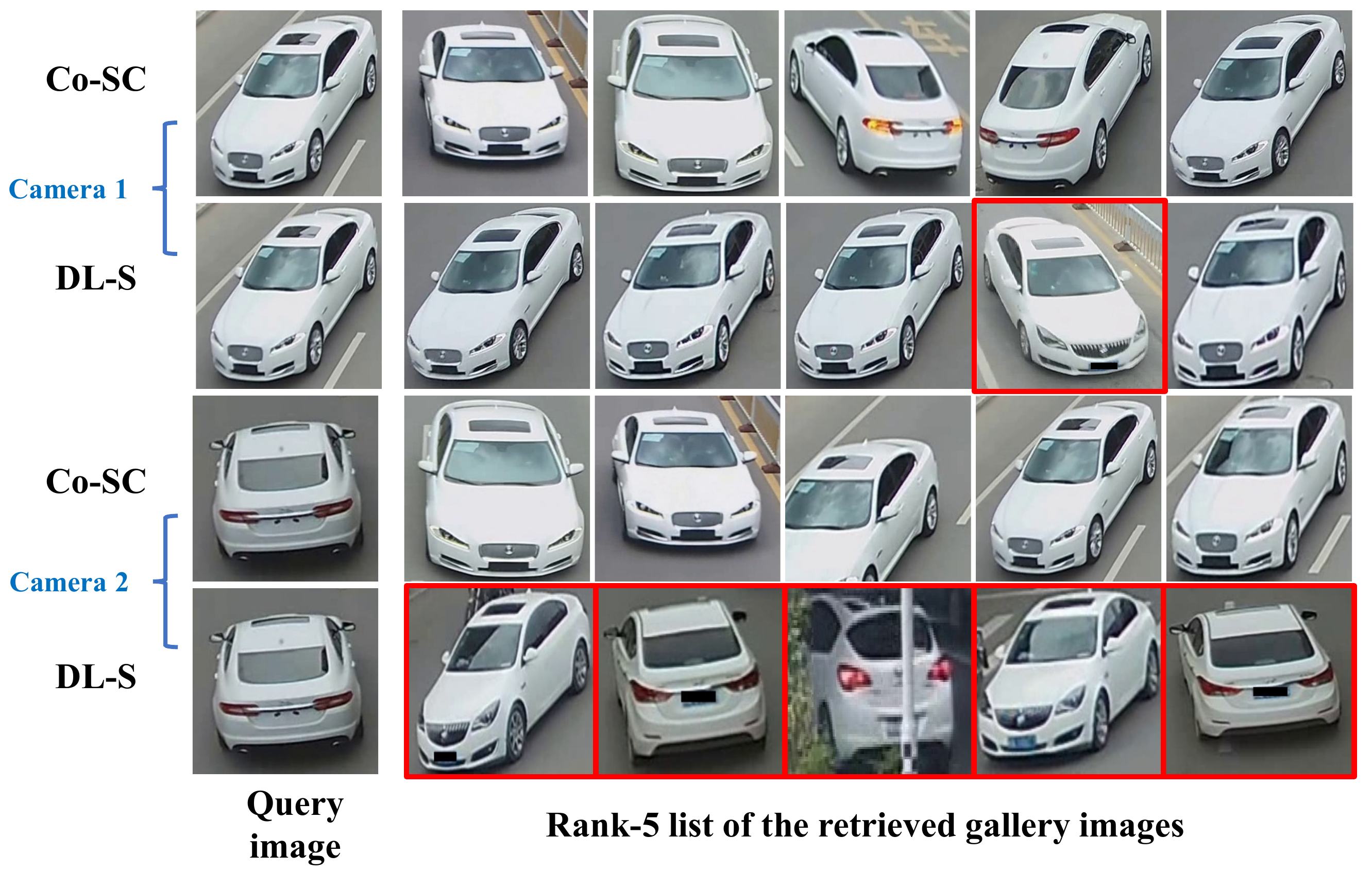}}\quad
	\subfloat[Visualization of feature maps.]{\label{Fig_visualization}\includegraphics[width=7.9cm]{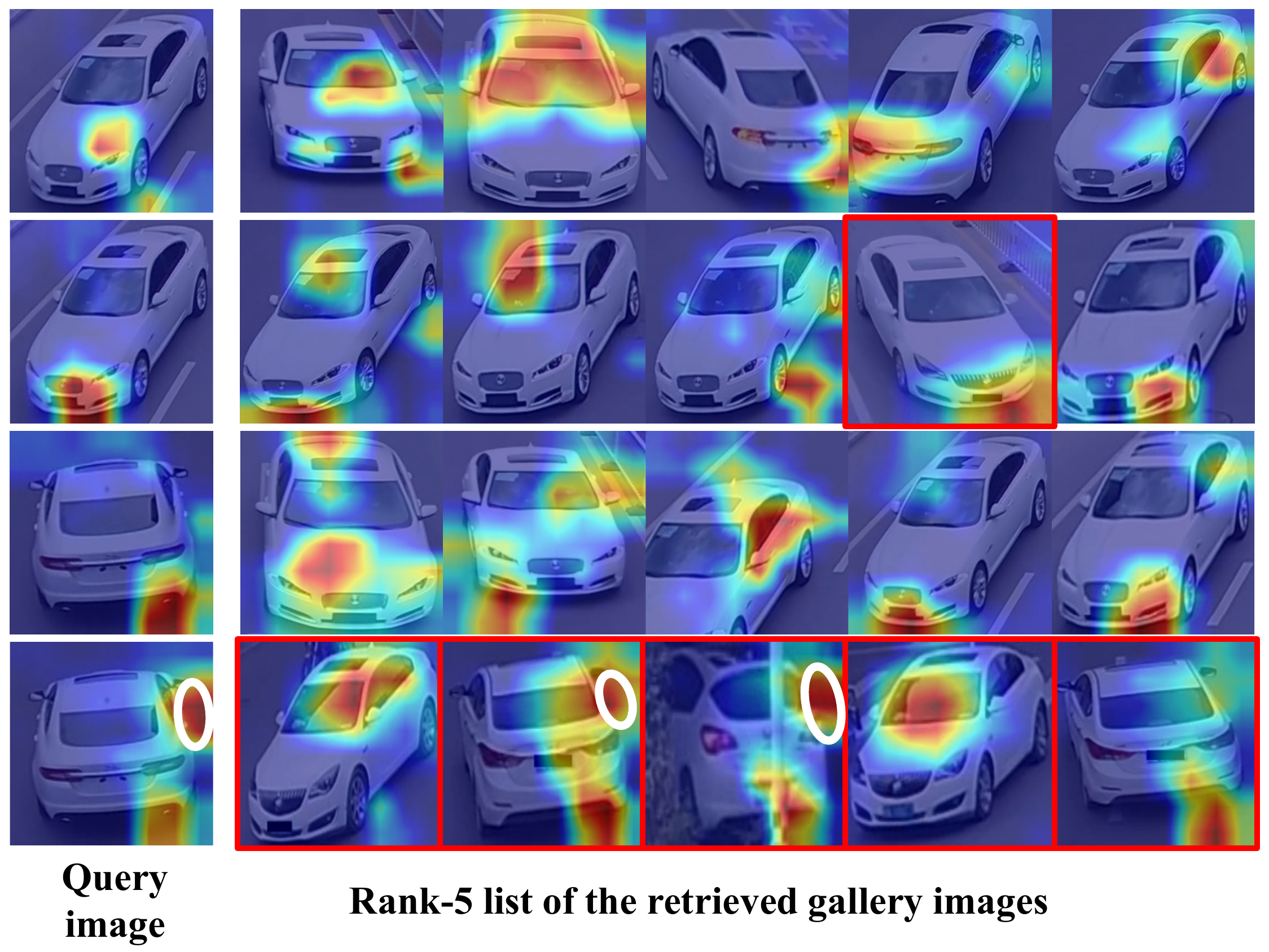}}\\	
	\caption{(a) Retrieval results of Co-SC and \textit{DL-S}.   {The red boxes indicate incorrect results, which are the retrieved gallery images of different identification with the query image.} (b) Grad-CAM visualization of feature maps. Warmer color indicates the contributing semantic features. White circles indicate distracting backgrounds.}
	\label{Fig_retrieval_visualization}
\end{figure*}

The parameters of Co-SC and DL-based baselines are given in Table~\ref{parameter setting of proposed}.
  {The common-used training configurations can be referred to~\cite{yimeng2022}.}

\begin{table}[htbp]
\centering
\caption{Parameter Settings of Co-SC and DL-based baselines}
\begin{center}
\renewcommand\arraystretch{1.3}
\begin{tabular}{|c|c|c|c|}
\hline
\textbf{Module} & \textbf{\thead{Layer\\Name}} & \textbf{\thead{Output\\Dimension}} &\textbf{Activation} \\
\cline{1-1} \cline{2-4}
\textbf{Semantic Encoder$^{\mathrm{a}}$} &  ResNet-50$^{\mathrm{d}}$  & $F$ &  \verb|\| \\
\hline
\multirow{2}{*}{\textbf{JSC Encoder}} &  FC with BN$^{\mathrm{e}}$ &  $2B$ &  Leaky ReLu  \\
\cline{2-4}
&  FC & $2B$ & Linear \\
\hline
\textbf{Identifier} &  FC  &  576 &  Softmax \\
\hline
\rowcolor{violet!10}
& CNN & $16B$  & None  \\
\cline{2-4}
\rowcolor{violet!10}
&  FC with BN  &  $F$ &  Leaky ReLu \\
\cline{2-4}
\rowcolor{violet!10}
\multirow{-3}{*}{\textbf{\thead{Cooperative\\JSC Decoder$^{\mathrm{b}}$}}} &  FC  &  $NF$ & Leaky ReLu \\
\hline
\rowcolor{violet!10}
\textbf{Fusion Module} &  CNN  & $F$   &  None \\
\hline
\rowcolor{violet!10}
\textbf{Gating Module} &  FC  &  2 &  Sigmoid \\
\hline
\rowcolor{cyan!8}
& CNN & $8B$  & None  \\
\cline{2-4}
\rowcolor{cyan!8}
&  FC with BN  &  $F$ &  Leaky ReLu \\
\cline{2-4}
\rowcolor{cyan!8}
\multirow{-3}{*}{\textbf{JSC Decoder$^{\mathrm{c}}$}} &  FC  &  $F$ &  Leaky ReLu \\
\hline
\multicolumn{4}{l}{$^{\mathrm{a}}$ White rows indicate the shared modules of DL-based methods.}\\
\multicolumn{4}{l}{$^{\mathrm{b}}$ Purple rows indicate the specific modules of Co-SC.}\\
\multicolumn{4}{l}{$^{\mathrm{c}}$ Blue rows indicate the specific module of DL-based baselines.}\\
\multicolumn{4}{l}{$^{\mathrm{d}}$ For ResNet-50, we set $F=2048$.}\\
\multicolumn{4}{l}{$^{\mathrm{e}}$ \textit{FC} indicates fully-connected layer. \textit{BN} indicates batch normalization.}\\
\end{tabular}
\label{parameter setting of proposed}
\end{center}
\end{table}

For simplicity, a two-user case is considered. Single-tap Rayleigh fading channels are adopted with channel coefficients following $\mathcal{CN}(0, 1)$. The power budget $P$ is set as $1$ for each user. The number of receiver antennas $M$ is $4$. {The number of transmitted complex symbols by DL-based methods $B$ is set as $16$ in all simulations.}

% {We provide some visualization results of identification in Fig.~\ref{Fig_retrieval_visualization}, 
Fig.~\ref{Fig_retrieval_visualization} provides visualization results of the identification task with the proposed Co-SC architecture and the DL-S baseline.
Fig.~\ref{Fig_retrieval_visualization}\subref{Fig_retrieval} shows the original query image provided by two cameras and the rank-5 list of the corresponding retrieved gallery images, where the incorrect results are marked with red boxes.
Fig.~\ref{Fig_retrieval_visualization}\subref{Fig_visualization} is the visualization of the contributing feature maps obtained by Grad-CAM~\cite{selvaraju2017grad}, where the warmer color indicates features with more significance.
Leveraging the correlation among users, Co-SC can retrieve more correct results by combining the informative semantic features, even when the query image is quite different from gallery images. 
  {
As shown in Fig.~\ref{Fig_retrieval_visualization}\subref{Fig_retrieval}, in the third line, the query image from camera 2 is the back of a vehicle, and Co-SC helps the server retrieve more gallery images corresponding to the correct vehicle, including images captured from the front view,
by incorporating the semantic feature of front-view image from camera 1.
% As shown in Fig.~\ref{Fig_retrieval}, the query image from camera 2 is the back of a car, and Co-SC helps the server accurately retrieve the gallery images captured from the front view
% by incorporating the semantic feature of front-view image from camera 1. 
In comparison, the non-cooperative method \textit{DL-S}
% for \textit{DL-S}, the communication and identification are non-cooperatively performed among users. As a result, \textit{DL-S} 
fails to distinguish the vehicles with similar looks, where all the retrieved gallery images for camera 2 are incorrect, as shown in the fourth line Fig.~\ref{Fig_retrieval_visualization}\subref{Fig_retrieval}.} 
% and is prone to be distracted by irrelevant features, such as backgrounds.
The further analysis shown in Fig.~\ref{Fig_retrieval_visualization}\subref{Fig_visualization} indicates that \textit{DL-S} is prone to be distracted by irrelevant features, such as backgrounds, due to the lack of cooperation between users.
In specific, as shown in Fig.~\ref{Fig_retrieval_visualization}\subref{Fig_visualization}, \textit{DL-S} puts more attention on the backgrounds of the query image and some gallery images, which are highlighted with white circles. This misleads the server to classify these images as the same identity due to the great similarity between the semantic features of the backgrounds, instead of cars.

\begin{figure}
\centering
\includegraphics[width=8cm]{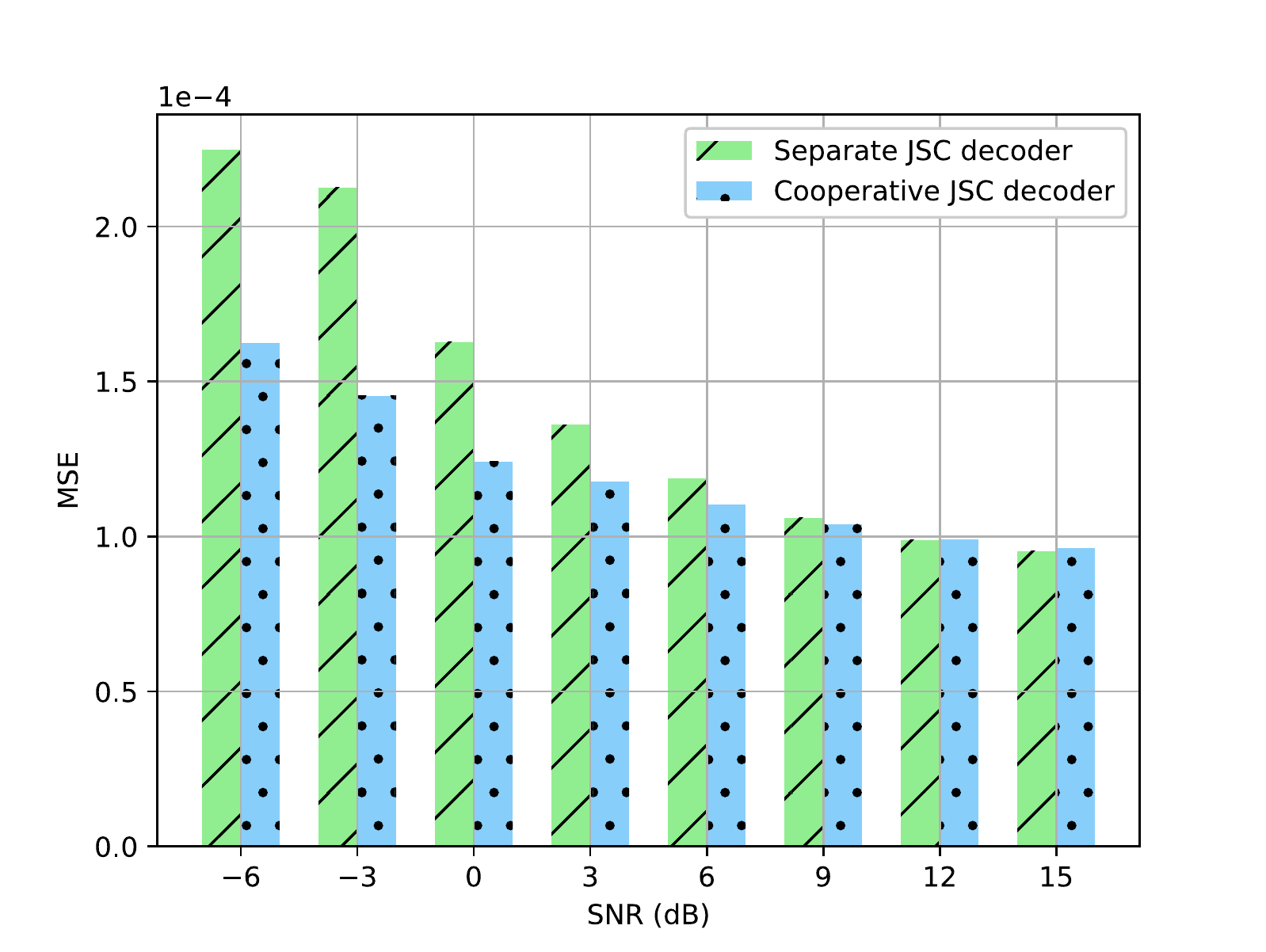}
\caption{The MSE of cooperative JSC decoder and separate JSC decoder.}\label{Fig_MSE}
\end{figure}

{Fig.~\ref{Fig_MSE} shows the MSE of the cooperative JSC decoder in Co-SC and the separate JSC decoder in \textit{DL-S}. It can be observed that the MSE of the two methods both decreases with the increase of signal-to-noise-ratio (SNR). The cooperative JSC decoder improves the recovery performance significantly in the low SNR regime and achieves similar performance with the separate decoder when SNR is high. In other words, the semantic-level correlation facilitates a more robust transmission by the proposed cooperative JSC coding scheme.}

\begin{figure}[tbp]
	\centering
	\subfloat[Rank-1 Accuracy of Different Methods]{\label{fig:a}\includegraphics[width=8cm]{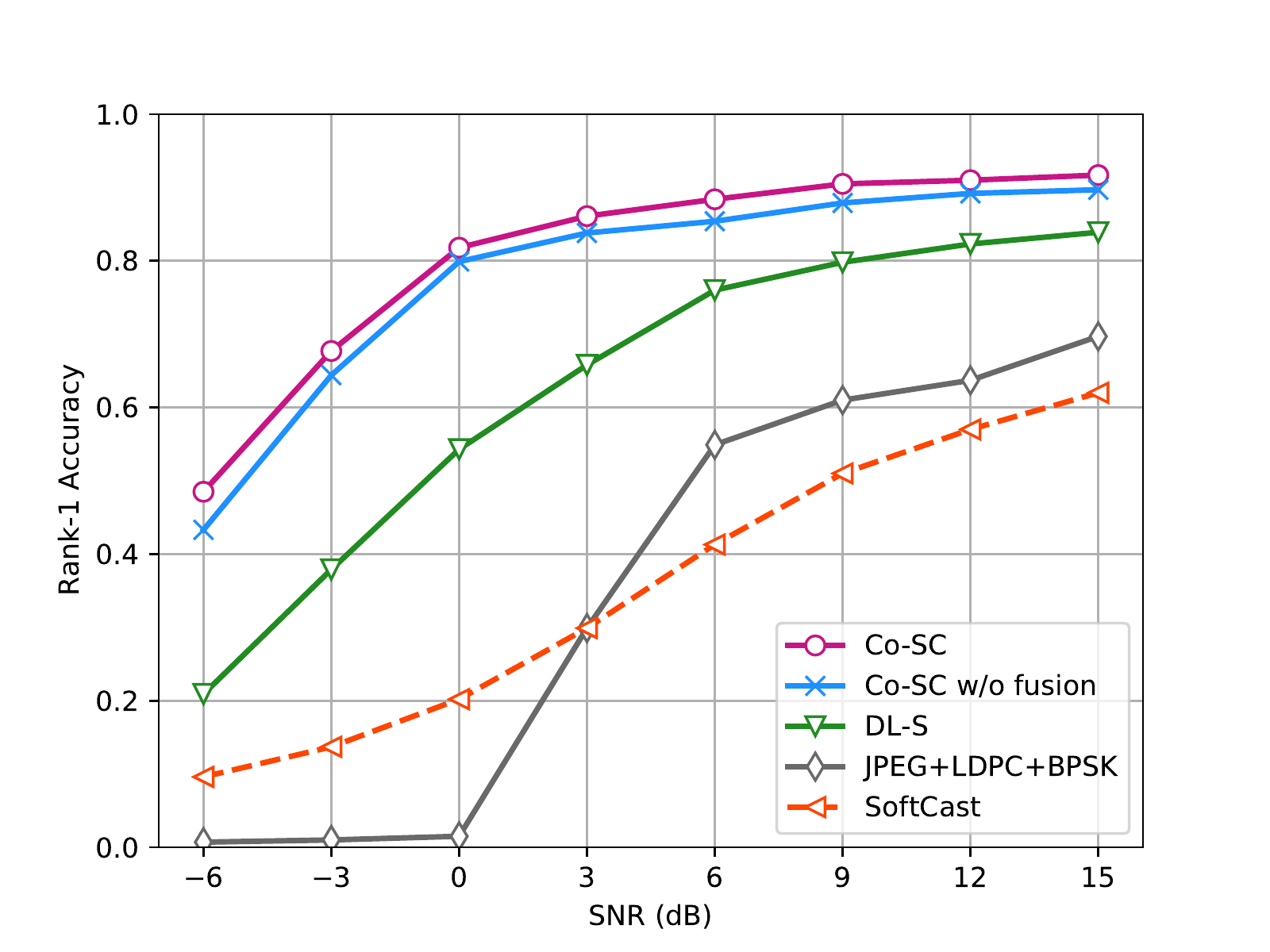}}\label{Fig_rank}\quad
	\subfloat[mAP of Different Methods]{\label{fig:b}\includegraphics[width=8cm]{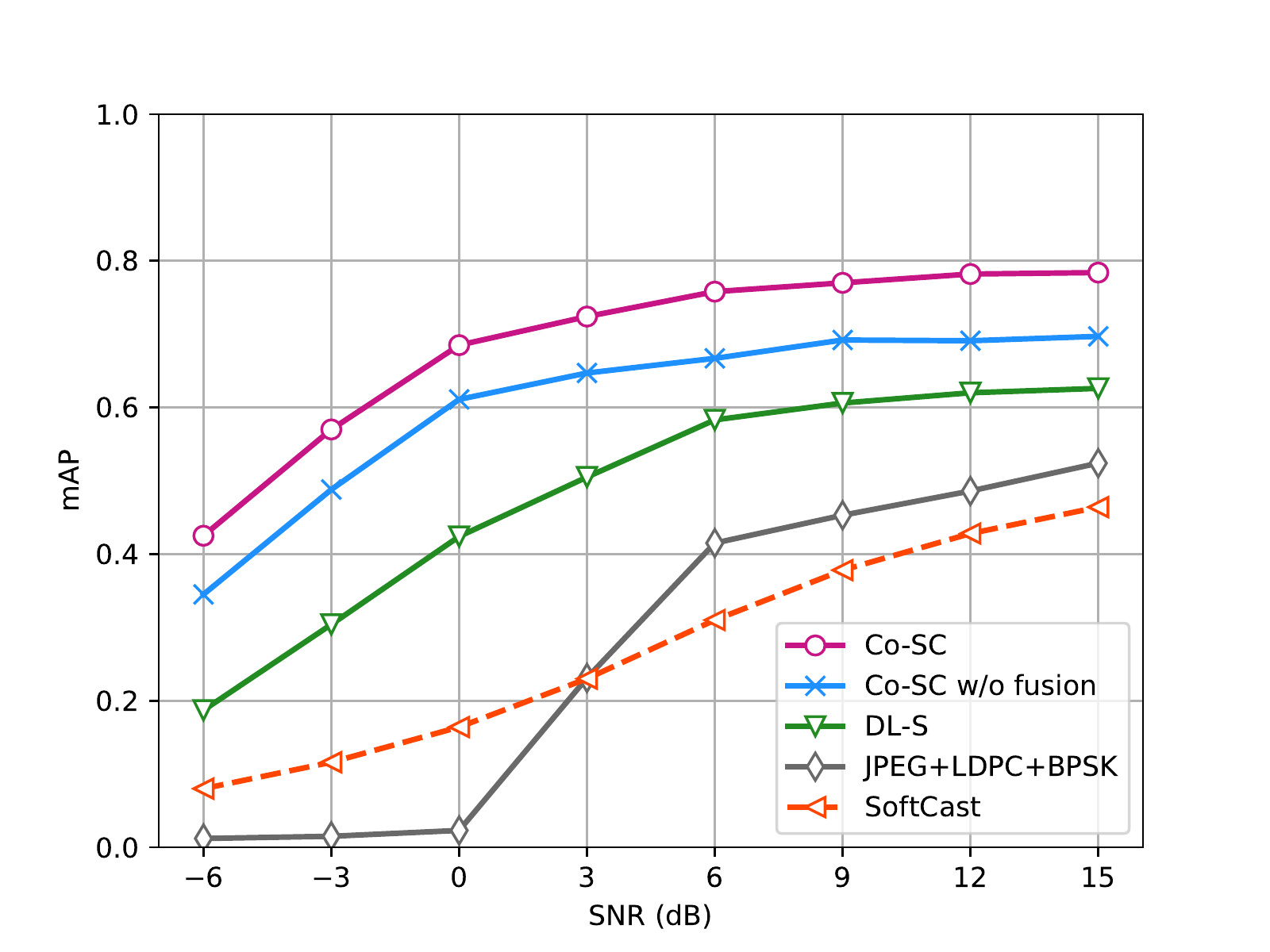}}\label{Fig_map}\\	
	\caption{  {Rank-1 accuracy and mAP comparison between Co-SC, \textit{Co-SC w/o fusion}, \textit{DL-S}, \textit{JPEG+LDPC+BPSK}, and SoftCast under Rayleigh channels.}}
	\label{Fig_rank_map_ray}
\end{figure}

The identification performance is presented in Fig~\ref{Fig_rank_map_ray}, where
DL-based semantic transmission methods with limited symbols all outperform the traditional \textit{JPEG+LDPC+BPSK} and SoftCast method.
  {Moreover, the average number of symbols used in the two traditional methods are about $3.1 \times 10^{5}$ and $2.6 \times 10^{5}$, respectively, much more than the proposed semantic-based method with 16 symbols, verifying the superiority of semantic communications in reducing data traffic.}
% Compared with \textit{DL-S}, \textit{Co-SC w/o fusion} improves the identification accuracy significantly, especially in low SNR regime, where the gap in terms of rank-$1$ accuracy and mAP is 26.5\% and 18.4\% at $-3$~dB, respectively.
Co-SC achieves the best performance among three semantic transmission methods, followed by \textit{Co-SC w/o fusion} and \textit{DL-S}.
At $-3$~dB, the gap of rank-$1$ accuracy between Co-SC and \textit{Co-SC w/o fusion}, \textit{DL-S} are 2.4\% and 28.9\%, respectively,
while in terms of mAP, which evaluates the identification performance in a global view, the gap are 8.3\% and 26.7\%, respectively.
% \begin{figure*}
% \centering
% \includegraphics[width=8.5cm]{visu3.pdf}
% \caption{Comparison of retrieval results. The red boxes indicate incorrect results.}\label{Fig_retrieval}
% \end{figure*}

% \begin{figure}
% \centering
% \includegraphics[width=8cm]{visualization.pdf}
% \caption{Visualization of feature maps.}\label{Fig_visualization}
% \end{figure}

\section{Conclusions and Outlook}
In this article, we have proposed a cooperative semantic-aware architecture for multi-user communications in IoV to reduce the data traffic significantly.
Such a revolution has achieved a transformation from traditional syntactic communications among users to semantic communications for IoV applications.
We have presented the main guidelines and principles of the architecture designed for cooperative semantic communications.
The proposed architecture is flexible enough to be adapted to different applications.
We have highlighted the advantages of the proposed architecture by implementing a case study.
Experimental results show that 1) conveying semantics in the source data requires less spectrum resources compared to the conventional syntactic symbol transmission; 2) correlations of semantics among different users leveraged by the cooperative JSC coding scheme achieves better semantic reconstruction performance without an extra communication overhead.
Consequently, the proposed architecture requires less radio resources to achieve better performance, easing the spectrum scarcity challenge in IoV.

This article is an initial work to present a view of conveying semantics in IoV for multi-user communications.
  {
With dedicated transformations, 
%	and auxiliary modules like edge knowledge base
	the proposed architecture is promising in serving more cooperative communication scenarios, like the communication from server to vehicular users.
}
Substantial further research is required in the following areas:
\begin{itemize}
    \item \textit{Knowledge base update}: The knowledge base of users and servers can evolve over time. How to model and keep track of the knowledge base variation for further improving the system performance is an essential open issue.
    \item \textit{Theoretical analysis of semantic-aware networks}: Semantic-aware networks heavily rely on artificial intelligence technologies to extract essential semantics for specific tasks. Due to the lack of mathematical formulation, the quantitative analyse, including semantic channel capacity, semantic distortion and the relationship between semantic channel capacity and syntactic channel capacity, are still missing.
    Developing a theoretical analysis methodology is vital.
    \item \textit{Semantic-aware security policy}: The security of semantic communication is more complex and should be closely related to the knowledge base sharing between users and servers. Developing an effective semantic-aware security policy is an essential direction for future research.
\end{itemize}

\section{Acknowledgments}
\noindent This work was supported in part by the National Natural Science Foundation of China under Grant 62293485, in part by the Fundamental Research Funds for the Central Universities under Grant 2022RC18, and in part by the China Scholarship Council. Fengyu Wang is the corresponding author of this article.

\bibliographystyle{IEEEtran}
\bibliography{IEEEabrv, iov,semantic,mypub, ML}

\begin{IEEEbiography}[{\includegraphics[width=1in,height=1.25in,clip,keepaspectratio]{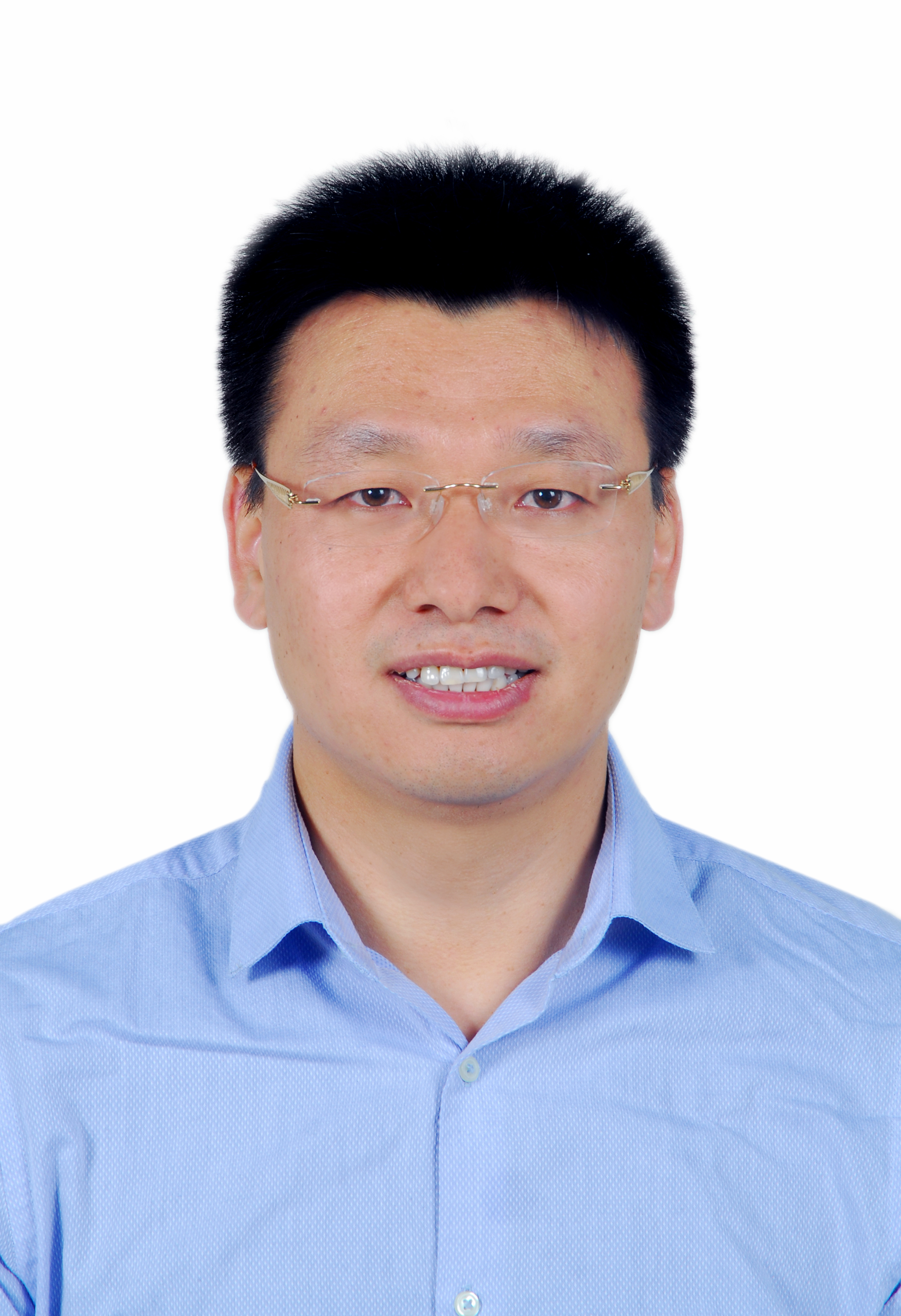}}]{Wenjun Xu} (wjxu@bupt.edu.cn) is a professor with the School of Artificial Intelligence, Key Lab of Universal Wireless Communications, Ministry of Education, Beijing University of Posts and Telecommunications, Beijing, China, and also with Peng Cheng Laboratory, Shenzhen, China. His research interests include AI-driven networks, semantic communications, UAV communications and networks, green communications and networking, and cognitive radio networks. He is serving as an Editor of China Communications. He is a Senior Member of the IEEE.
\end{IEEEbiography}

\begin{IEEEbiography}[{\includegraphics[width=1in,height=1.25in,clip,keepaspectratio]{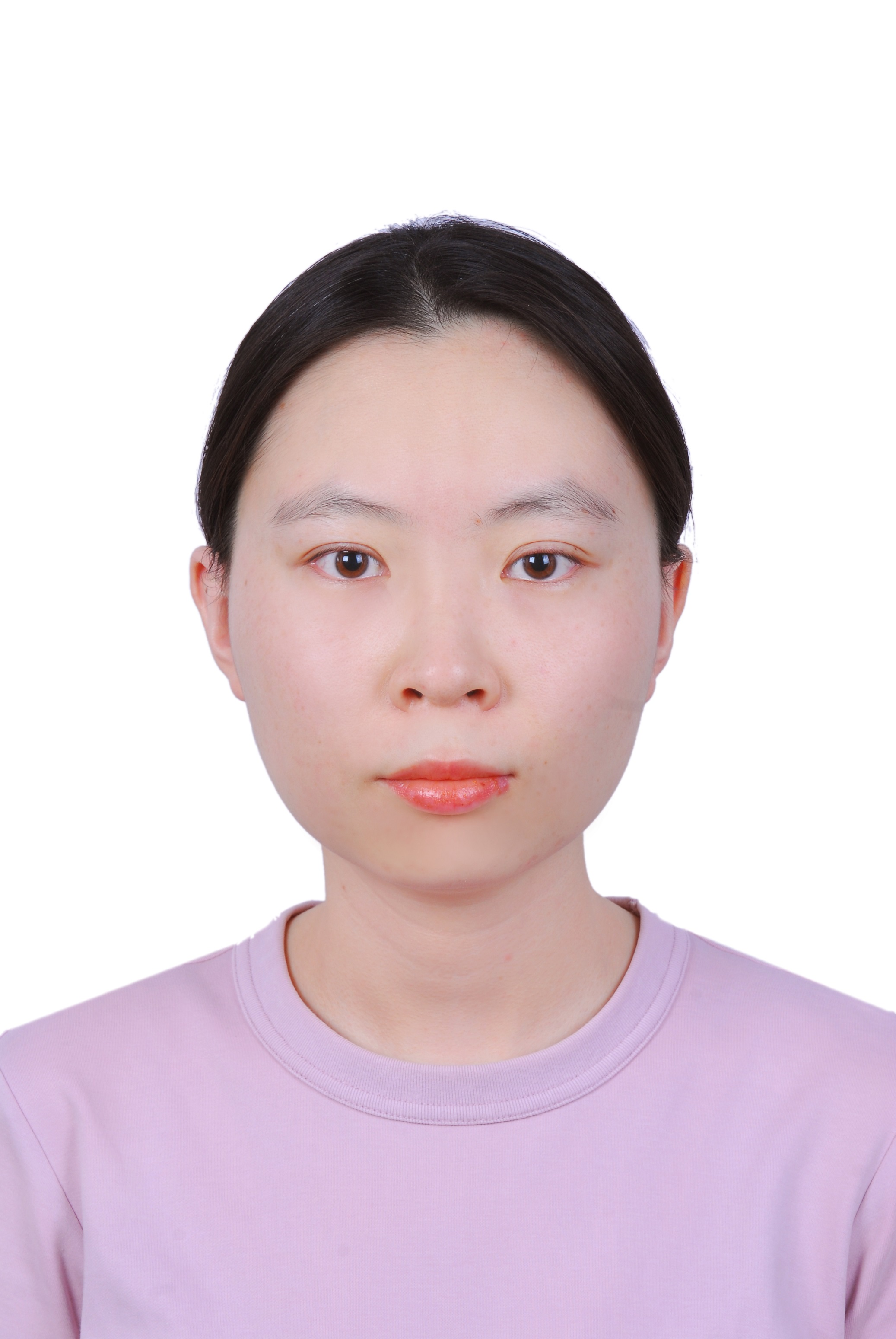}}]{Yimeng Zhang} (yimengzhang@bupt.edu.cn) is currently pursuing her Ph.D. degree at the School of Artificial Intelligence, Key Lab of Universal Wireless Communications, Ministry of Education, Beijing University of Posts and Telecommunications, Beijing, China.
Her current research interests include semantic communications, intelligent resource allocation in emerging wireless applications. She is a Graduate Student Member of the IEEE.

\end{IEEEbiography}

\begin{IEEEbiography}[{\includegraphics[width=1in,height=1.25in,clip,keepaspectratio]{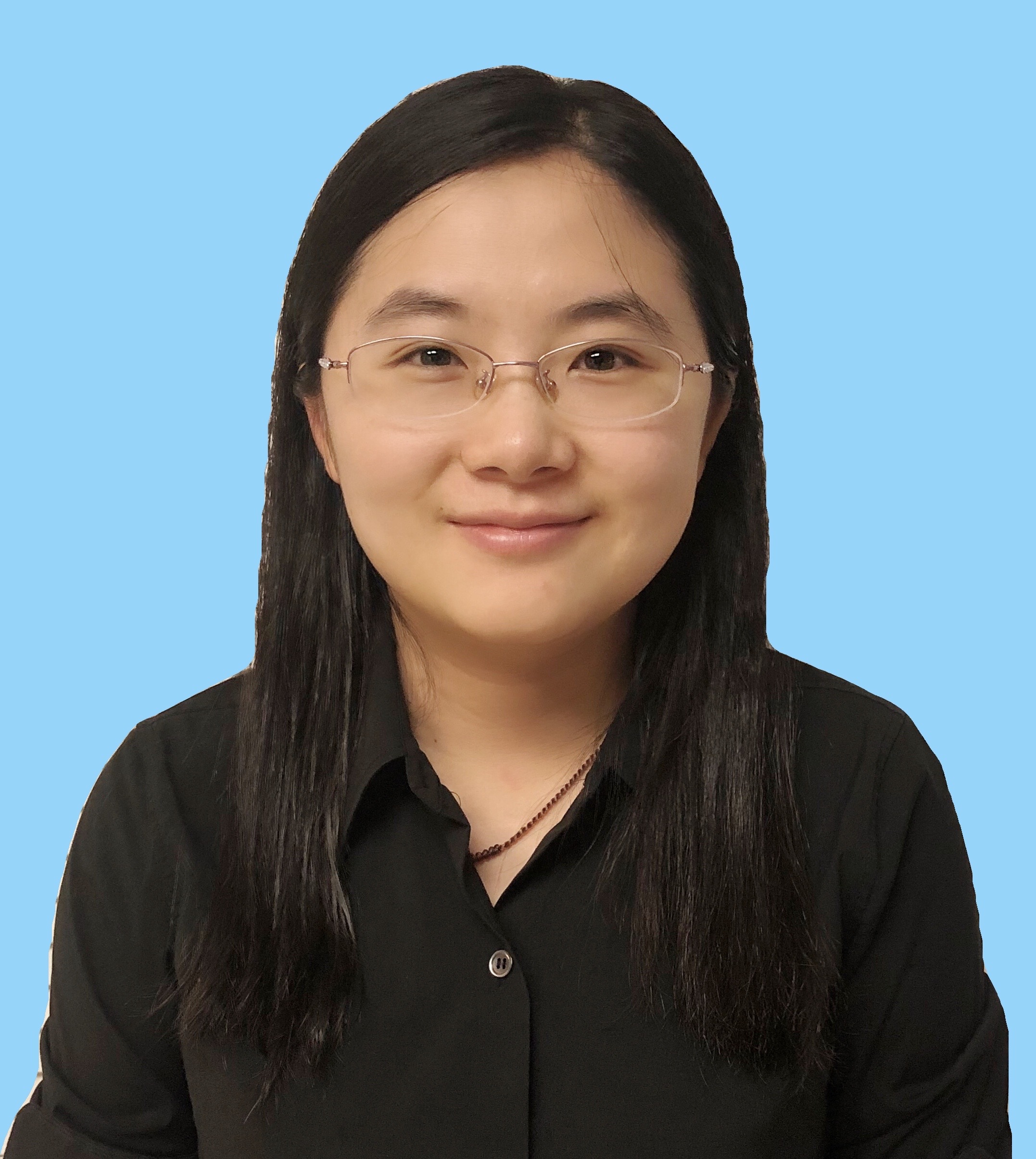}}]{Fengyu Wang} (fengyu.wang@bupt.edu.cn) is currently a lecturer with the School of Artificial Intelligence, Beijing University of Posts and Telecommunications, Beijing, China.
Her current research interests include Integrated sensing and communications (ISAC), semantic communications, wireless sensing and statistical signal processing. She is a Member of the IEEE.
\end{IEEEbiography}

\begin{IEEEbiography}[{\includegraphics[width=1in,height=1.25in,clip,keepaspectratio]{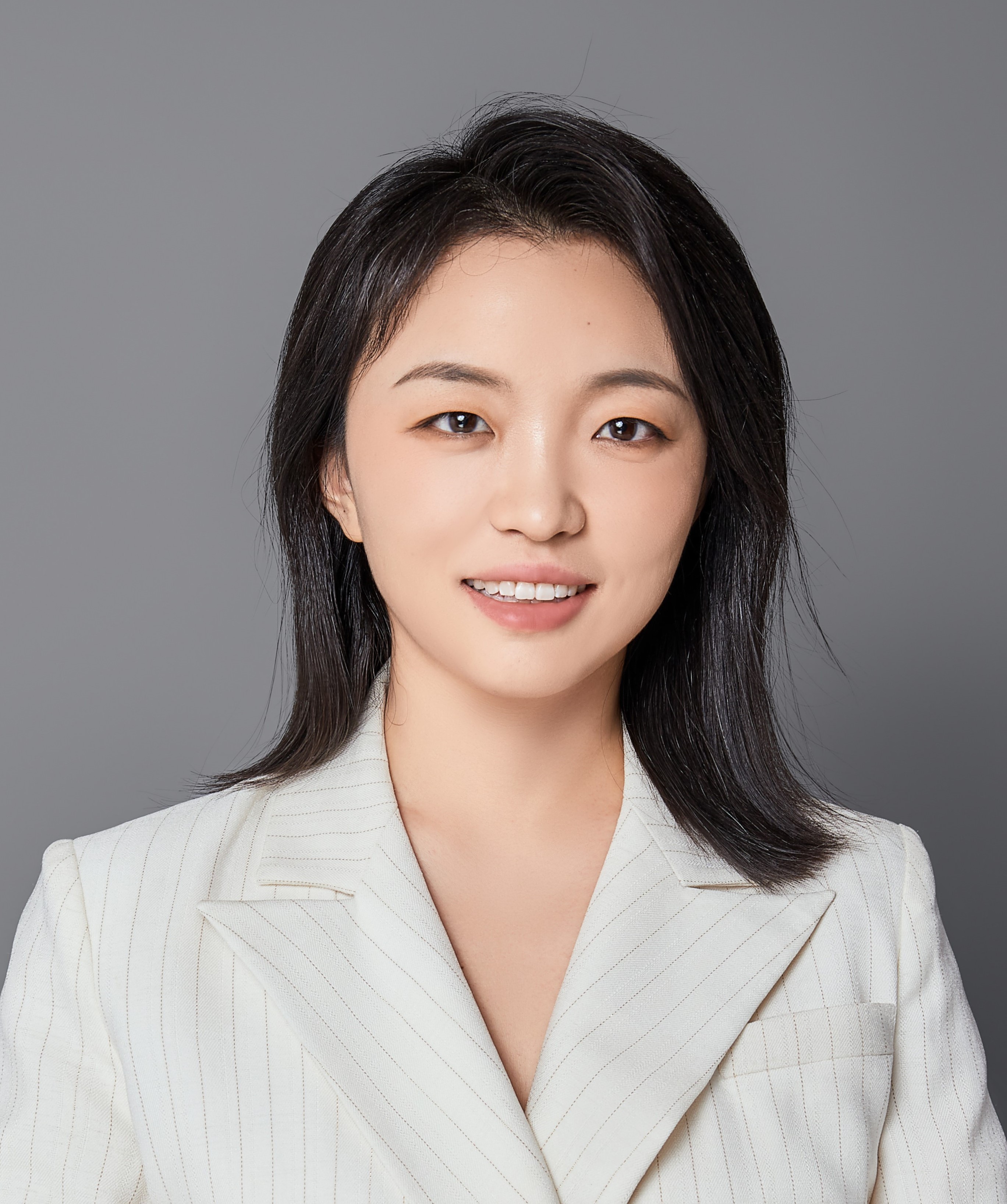}}]{Zhijin Qin} (qinzhijin@tsinghua.edu.cn) is currently an Associate Professor with the Department of Electronic Engineering, Tsinghua University, Beijing, China. Her research interest is semantic communications. She is serving as an associate editor of IEEE Transactions on Communications, IEEE Transactions on Cognitive Networking, and IEEE Communications Letters. She has received several awards from IEEE Communications Society and IEEE Signal Processing Society. She is a Senior Member of the IEEE.
\end{IEEEbiography}

\begin{IEEEbiography}[{\includegraphics[width=1in,height=1.25in,clip,keepaspectratio]{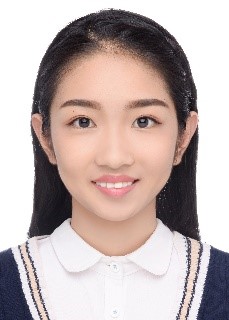}}]{Chenyao Liu} (liuchenyao@bupt.edu.cn) is currently pursuing her Ph.D. degree at the School of Artificial Intelligence, Key Lab of Universal Wireless Communications, Ministry of Education, Beijing University of Posts and Telecommunications, Beijing, China. Her research interests include semantic communications, video coding and machine learning.
\end{IEEEbiography}

\begin{IEEEbiography}[{\includegraphics[width=1in,height=1.25in,clip,keepaspectratio]{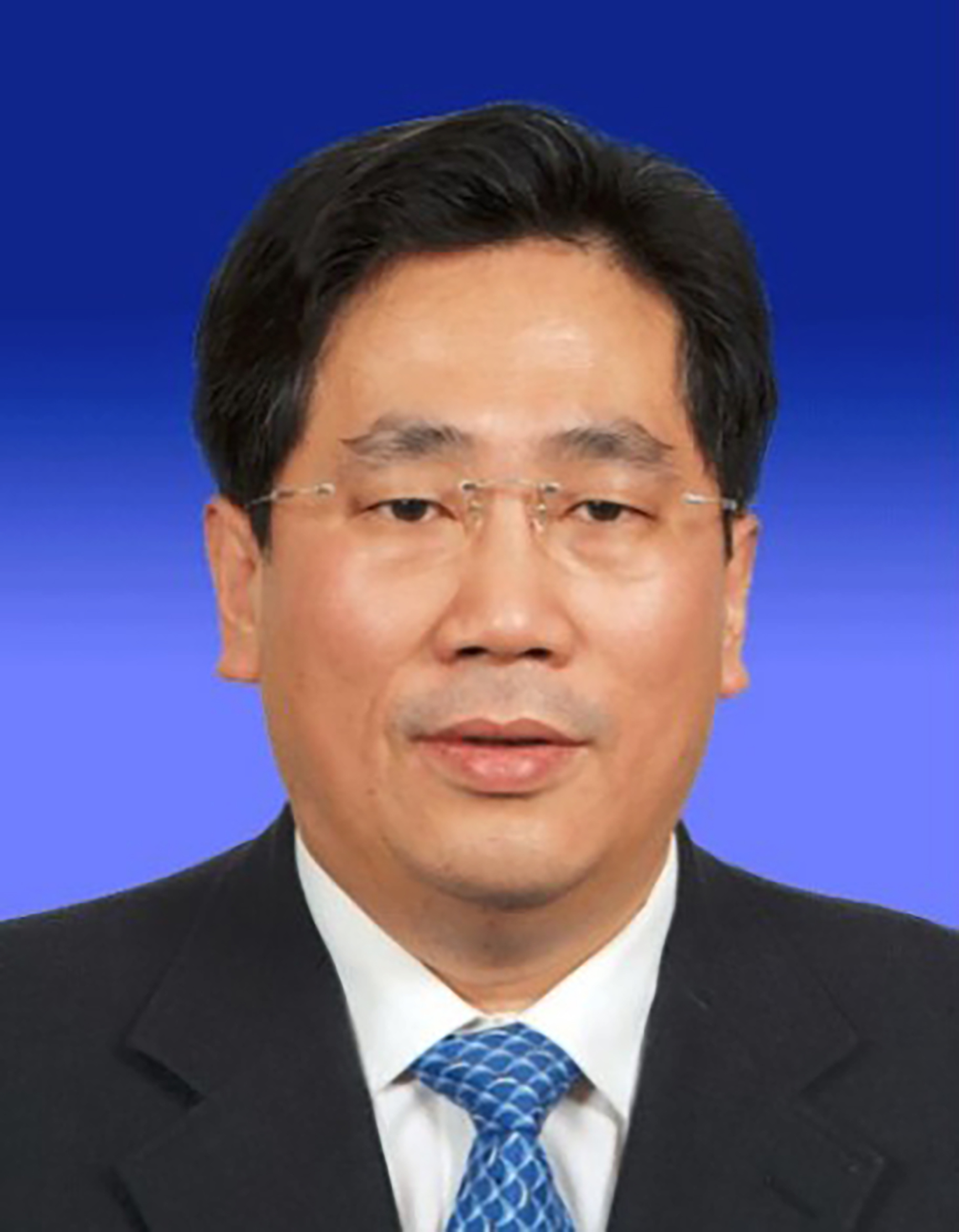}}]{Ping Zhang} (pzhang@bupt.edu.cn) is currently a professor with the School of Information and Communication Engineering at Beijing University of Posts and Telecommunications, the director of the State Key Laboratory of Networking and Switching Technology, and also with the Department of Broadband Communication, Peng Cheng Laboratory, Shenzhen, China. His current research interests mainly focus on wireless communications. He is an Academician of the Chinese Academy of Engineering and a Fellow of IEEE.
\end{IEEEbiography}

\end{document}